# Enhanced CBF Packet Filtering Method to Detect DDoS Attack in Cloud Computing Environment

Priyanka Negi[1], Anupama Mishra[2] and B. B. Gupta[3]

[1,2]Graphic Era University
Dehradun, Uttarakhand 248001, India
[1]priyankanegi25dec@gmail.com
[2]tiwari.anupama@gmail.com

[3] RSCOE, University of Pune, India
[3]gupta.brij@gmail.com

**Abstract**
Tremendous and extraordinary growths in the field of internet, intranet, extranet and its users have developed an innovative era of great global competition and contention. Denial of service attack by multiple nodes is accomplished of disturbing the services of rival servers. The attack can be for multiple reasons. So it is a major threat for cloud environment. Due to low effectiveness and large storage conventional defending approaches cannot be easily applied in cloud security. The effects of various attacks can decrease the influence of a cloud. So, in view of this challenge task, this paper aims at enhancing a proposed method for cloud security. We propose a modification to the confidence Based Filtering method (CBF) which is investigated for cloud computing environment based on correlation pattern to mitigate DDoS attacks on Cloud. The modification introduces nominal additional bandwidth and tries to increase the processing speed of the victim initiated server.

## 1. Introduction

Cloud computing basically referred to applications and service that are offered over the internet from the data centers all over the world. So it is the delivery of computing as a service rather than a product, whereby shared resources, software, and information are provided to computers and other devices as a utility over a network. NIST defines cloud computing as "Cloud computing is a model for enabling convenient, on-demand network access to a shared pool of configurable computing resources (e.g., networks, servers, storage, applications, and services) that can be rapidly provisioned and released with minimal management effort or service provider interaction."[1,2].
The cloud model is composed of five essential characteristics, three service models, and four deployment models. The Essential Characteristics are On-demand self-service, broad network access, Resource pooling, rapid elasticity, Measured Service. Services of cloud computing is offered on the basis of three models i.e Infrastructure as a service (IaaS), Platform as a service (PaaS) and Software as a service (SaaS). As it provide large amount of resources online, so it is facing with several security problems.

The security issues on cloud computing primarily focus on data safety, data privacy, data confidentiality and network security. Considering malicious intruders, there are many kinds of possible attacks, such as: Wrapping attack, Malware-Injection attack, Flooding attack, Browser attack. A Wrapping attack is done by duplication of the user account and password in the log-in phase so that the SOAP (Simple Object Access Protocol) messages that are exchanged during the setup phase between the Web browser and server are affected by the attackers. In a Malware-Injection attack, the attacker creates a normal operation, such as to delete the user, and embeds in it another command, such as setting the administration rights. So, when the user request is passed to the server, rather than the server executing the command as if it were deleting a user account, it actually discloses a user account to the attacker [3].

A Flooding attack [7,8] occurs when an attacker generates bogus data, which could be resource requests or some type of code to be run in the application of a legitimate user, engaging the server's CPU, memory and all other devices to compute the malware requests. The servers finally end up reaching their maximum capacity, and thereby offload to another server, which results in flooding. A Browser attack is committed by subverting the signature and encryption during the translation of SOAP messages in between the web browser and web server, causing the browser to consider an adversary as a legitimate user and process all requests communicating with web server [4].

In this article we present a modification to the confidence Based Packet filtering Technique [12-14] to address the problem of database and processing speed on victim initiated side. In the proposed scheme to handle the database problem the option filed of IP header is used to store the information of packet.

The rest of the article is structured in the following way: all the literature survey on flooding attacks and packet filtering techniques is discussed in Section II and III; the basic CBF method and its limitation, modifications proposed for the basic CBF method to alleviate the problem associated with database and processing speed are presented in Section IV; the discussion and analysis of proposed method also are described in Section IV; finally, we conclude in Section V.

## 2. Flooding attack

Flooding attack, basically consist of an attacker sending a huge amount of nonsense requests to a certain service, which is providing various services under cloud. As each of these requests has to be processed by the service implementation in order to determine its invalidity, this causes a certain amount of workload per attack request, which in the case of a flood of requests usually would cause a Denial of Service to the server hardware.

## 3. Current status of Related Research

Cloud computing refers to the delivery of computing and storage capacity as a service to a heterogeneous community of end-recipients. It relies on sharing of resources to achieve coherence and economies of scale similar to a utility over a network. As discussed in [5] it has the potential of providing dramatically scalable and virtualized resources, bandwidth, software and hardware on demand to consumers. The top of this paragraph illustrates a sub-subheading.

In this area, a number of approaches that have already been proposed are: Packet Score filtering method generates value distributions of some attributes in the TCP and IP headers, and then uses Bayes' Theorem to score packets [9]. But in this method its scoring and discarding are related to attack intensity, so it is not suitable for handling large amount of attack traffic. As it has some costly operations in scoring, so it leads to low process efficiency in real-time filtering.

ALPi method is an improvement of Packet Score [10]. It uses two schemes LB and AV which uses leaky buckets and value variances of attributes and is evaluated by comparison with Packet Score.

Hop-Count Filtering (HCF) method [11] uses the relationship of source IP address and TTL value to carry out filtering. After building an IP to hop-count mapping, it can detect and discard spoofed IP packets. The limitation of this method is that it is vulnerable to distributed attacks because of its assumption about spoofed IP traffic.

CBF (Confidence-Based Filtering) method [12] is based on mining the correlation patterns, which refer to some simultaneously appeared characteristics in the legitimate packets. These patterns are mainly in network and transport layer. But in this method no fixed number of single attributes is defined that has to be selected. Apart from this problem a database is also maintained at the server side which uses the 3-dimensional array storing strategy due to which the processing speed of the server is slow down.

Table 1: KEY TERMS APPEARED IN THIS PAPER [12]

| Terms | Description |
| --- | --- |
| N | The number of attributes under consideration in the method |
| $A_i$ | The i-th attribute in the packet,($1 \leq i \leq n$) |
| $M_i$ | The number of values which $A_i$ can have |
| $a_{i,j}$ | The j-th value of attribute $A_i$,( $1 \leq j \leq m_i$) |
| T | A time interval in packet flows |
| $N_n$ | The total number of packets in the packet flow in one time interval t |
| $N(A_i = a_{i,j})$ | The number of packets whose attributes $A_i$ has value $a_{ij}$ in the packet flow in one time interval t |
| $N(A_r = a_{r,x}, A_s = a_{s,y})$ | The number of packets whose attribute $A_r$ has value $a_{r,x}$, attribute $A_s$ has value $a_{s,y}$ in this packet flow in one time interval t |
| P | A packet in the packet flows |
| p(i) | The value of attribute $A_i$ in packet p |

## 4. Enhanced CBF Packet Filtering Method

In the proposed work we will be modifying the CBF Packet filtering method so that utilization of storage at the victim side is reduced and the processing speed of the sever will be increased. The method will be based on Correlation Pattern.

| Version | IHL | Type of | Total |
|---------|-----|---------|-------|
| Identification | | Flags | Fragment Offset |
| Time to Live | | Protocol | Header |
| Source Address ||||
| Destination Address ||||
| Options ||||

0--------------------------------------------------15--------31

Fig. 1 Old IPV4 HEADER

But as we have to store the confidence value in the optional field of the IPv4 header, we have to add one 32 bit word in IP Header and to recognize that part of header we have to add 1 in the old IHL value of IPv4 old header. And so the new IPv4 header will be:

### 4.1 Confidence Value

The concept of confidence [12] reflects how much trust we can put on a correlation pattern between an attribute pair.

### 4.2 Confidence

Confidence is the frequency of appearances of attributes in the packet flows. The confidence for single attributes and attribute pairs are calculated as:

*Confidence for single attributes [12]:*

$$\text{Conf}(A_i = a_{i,j}) = \frac{N(A_i = a_{i,j})}{N_n}$$

where $i = 1, 2, 3, \ldots, n$, $j = 1, 2, 3, \ldots, m_i$.

*Confidence for attribute pairs [12]:*

$$\text{Conf}(A_{i1} = a_{i1,j1}, A_{i2} = a_{i2,j2}) = \frac{N(A_{i1} = a_{i1,j1}, A_{i2} = a_{i2,j2})}{N_n}$$

where $i1 = 1, 2, 3, \ldots, n$, $i2 = 1, 2, 3, \ldots, n$, $j1 = 1, 2, 3, \ldots, m_1$, $j2 = 1, 2, 3, \ldots, m_2$

The more times an attribute pair appears in the legitimate packet flows, the higher confidence value of this pair is.

So, with confidence values of attribute value pairs, the legitimacy criteria of a packet are defined.

### 4.3 Proposed Methodology

In the enhanced confidence-based filtering method legitimate packet is the one whose confidence based filtering value is above the discarding threshold. So, those packets with scores lower than the discarding threshold are regarded as attack ones. This method is based on two periods: attack period and Non attack period.

#### 4.3.1 Algorithm

```
Step1: Set the initial values to the required attributes
Step2: Declaring the Period whether it's attacking Period
or Non attacking Period
Step3: If   period is Non Attacking then
           Calculate the Confidence Value
               If   NP =NULL then
                   Update NP with the confidence value
               Else
               If Confidence Value< Value in NP   then
                   Update the NP and attach in packets
                   IHL=IHL+1
                   Option Field=Confidence Value
                   Accept the Packet
               Else
                   Attach in packets
                   IHL=IHL+1
                   Option Field=Confidence Value
                   Accept the Packet
               End if
           End if
        Else
          Set the Confidence value in the NP as discarding
threshold
Calculate the Confidence Value of packet
If Confidence Value (packet) < Discarding threshold then
        Discard the packet
    Else
        Accept the Packet
End if
End if
```

| Version | IHL+1 | Type of | Total |
|---------|-------|---------|-------|
| Identification | | Flags | Fragment Offset |
| Time to Live | | Protocol | Header |
| Source Address ||||
| Destination Address ||||
| Options ||||

0--------------------------------------------------15-------------31

Fig. 2 New IPV4 HEADER

### 4.3.2 Flow Chart

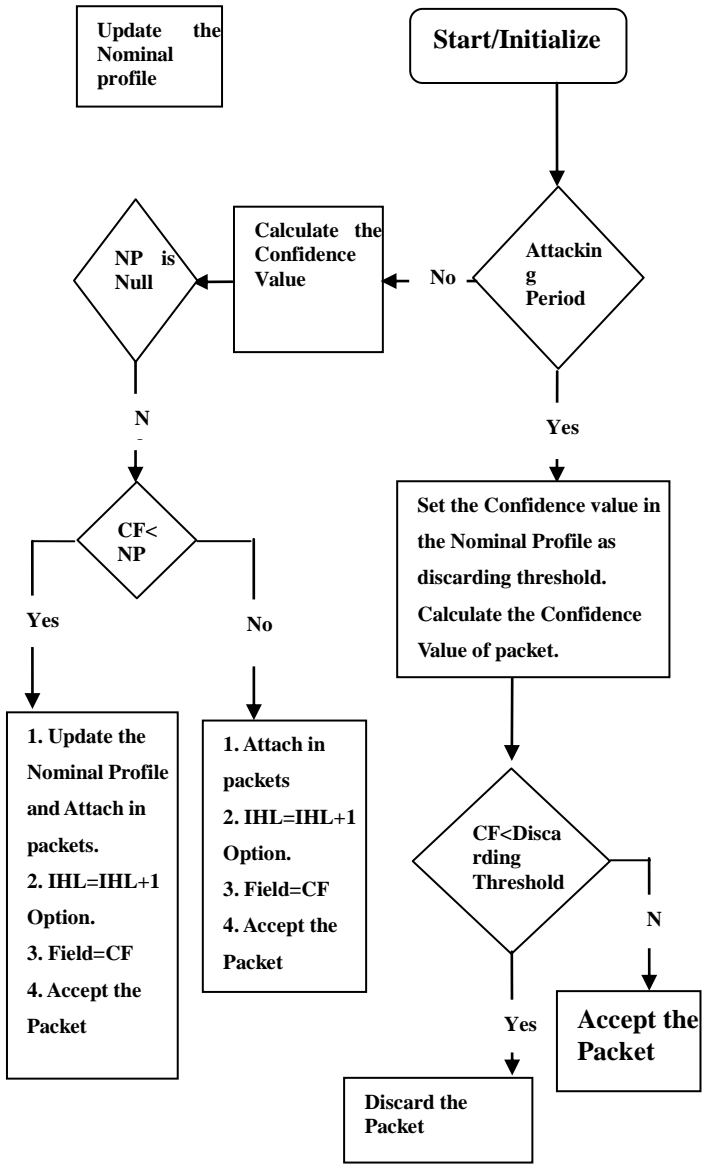

Fig. 3 Flowchart for Enhanced CBF Method

At attack period we will first set the confidence value stored in the nominal profile as the discarding threshold. Then will calculate the confidence value of the packet that comes in the attack period. If the confidence value of the packet is less than the discarding threshold value than we will discard the packet else if the confidence value of the packet is greater than the discarding threshold value then will accept the packet.

## 5. Conclusion and Scope for Future Work

The key concept of Enhanced CBF is based on correlation pattern, which is the co-appearance of attribute pairs. We introduced confidence to represent the distribution of attribute value pairs and then reduce the overhead of the sever by calculating the confidence value of the packet at the packet header itself and then storing the value in the optional field of the IPV4 packet header and at the same time updating the nominal profile variable if the value stored in the nominal profile is greater than the confidence value of the packet. Since the confidence reflects the frequency of appearances of the attribute value pairs, packets with more attribute value pairs of higher confidence will get higher score, which means more legitimate in this method. In the future we will try to use a set or group of single attributes so that to identify the correlation pattern will be more complex. In future we will try to simulate this enhanced CBF method on cloud simulator network security in cloud environment.

### Acknowledgments

This research is supported by Graphic Era University, Dehradun, India. Authors would like to thanks anonymous reviewers for their valuable suggestions and reviews.

**Priyanka Negi** received her Master in Computer Science and Engineering in 2012. She has published 5 research papers in International Journals and Conferences of high repute. Her research interest includes Cloud Computing and Security attacks.

**Anupama Mishra** received her Master in Computer Science and Engineering in 2010. Currently, she is a PhD candidate in Galgotias University, Greater noida, India. She has published 5 research papers in International Journals and Conferences of high repute. Her research interest includes Information Security, intrusion detection, Cloud Computing and Security attacks.

**B. B. Gupta** received PhD degree from Indian Institute of Technology Roorkee, India. Prior to that, he has obtained Bachelor of Engineering degree in Information Technology from the Rajasthan University, Jaipur, India. In 2009, he was selected for Canadian Commonwealth Scholarship and awarded by Government of Canada award. He has published more than 40 research papers in International Journals and Conferences of high repute. Dr Gupta is also holding position of editor of various International Journals and magazines of high repute. He also worked as a post doctoral research fellow in University of New Brunswick, Canada. His research interest includes Information security, Cyber Security, Intrusion detection, Computer networks and Phishing.